\documentclass[fleqn,12pt,twoside]{article}
\usepackage{espcrc1}

\usepackage{graphicx}

\newcommand{\rootmax}{$\sqrt{s_{NN}} = 200$ GeV}

\newcommand{\rootsps}{$\sqrt{s_{NN}} = 17$ GeV}

\title{Rapidity Dependence of Net--protons at \rootmax}

\author{P. Christiansen\address[NBI]
  {Niels Bohr Institute, University of Copenhagen, Denmark}%
  , for the BRAHMS collaboration}

\begin{document}

\maketitle

\begin{abstract}
  The BRAHMS collaboration has measured inclusive proton and
  anti--proton spectra at several rapidities ($0 \leq y \leq 3$) for Au+Au
  collisions at \rootmax. Rapidity densities of protons, anti--protons,
  and net--protons [$N(p)-N(\bar{p})$] are presented as a function of
  rapidity.
\end{abstract}

\section{INTRODUCTION}
\label{sec:intro}

In his article from 1983~\cite{bjorken83}, Bjorken proposed a scenario
for the evolution of the central rapidity region in ultra relativistic
heavy ion collisions. Two of his assumptions were motivated by what
had been observed in nucleon-nucleon collisions~\cite{proton}.  First,
the collision has to be transparent, meaning that the net--baryon
[$N(B)-N(\bar{B})$] content ends up at forward rapidities after the
collision.  The second assumption is the existence of a central
rapidity plateau with uniform particle production i.e., the invariant
yields $dN/dy$ are constant as a function of rapidity (boost
invariance). These assumptions allow this region to be described as
many identical thermal sources with different longitudinal velocities
undergoing only 1--dimensional longitudinal hydrodynamical evolution.
In this scenario the initial energy density can be related in a simple
way to the final particle production (used in fx.~\cite{ito02}).

At AGS and SPS where the degree of transparency is low, the original
baryons are stopped in the initial collisions and the mid--rapidity
region contains most of the net--baryons. The measured ratio of
anti--protons to protons at the Relativistic Heavy Ion Collider (RHIC)
at \rootmax, $N(\bar{p})/N(p) \approx 0.75$
\hspace{0.1cm}($y=0$) \cite{phobos02,bratio02}, indicates that particle
production in the central rapidity region is dominated by pair
production.

The BRAHMS experiment at RHIC consists of two magnetic spectrometers
with small solid angle coverage. The Mid Rapidity Spectrometer (MRS)
($30^\circ < \theta < 95^\circ$) uses Time-Of-Flight (TOF) for
particle identification (PID). In the Forward Spectrometer (FS)
($2.3^\circ < \theta < 30^\circ$) only the Ring Imaging Cherenkov
(RICH) is used in this analysis. The experiment is described in full
detail in~\cite{brahms}. By combining many different magnet and angle
settings, BRAHMS has a wide rapidity coverage for pions, kaons, and
protons, that allows the study of boost invariance and transparency at
RHIC.

\section{RESULTS}
\label{sec:results}

\begin{figure}[htbp]
  \vspace{-5mm}
  \begin{center}
    \includegraphics[keepaspectratio, width=0.75\columnwidth]
    {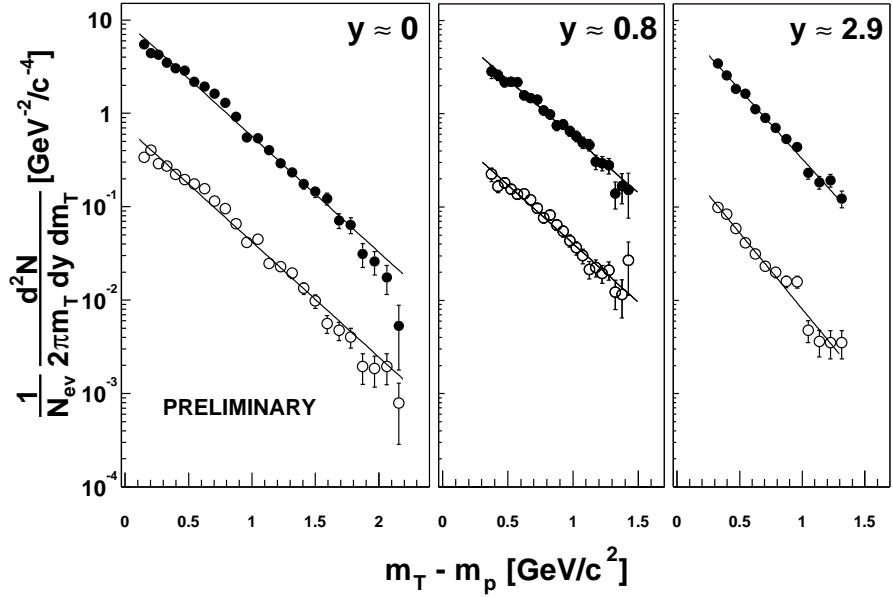}
    \vspace{-10mm}
    \caption{Transverse mass spectra for proton ($\bullet$) and
      anti--protons ($\circ$). The anti--proton spectra have been
      divided by a factor of 10 for clarity. Error bars are
      statistical only. The spectra have been fitted with
      Eq.~\ref{eq:fitfunc}.}
    \label{fig:spectra}
  \end{center}
  \vspace{-10mm}
\end{figure}

For all the data presented here, an event centrality cut of 0--10\%
($b \leq 4.4$ fm) was applied.  In the MRS, protons and anti--protons are
identified using cuts in the mass squared $m^2$ calculated from TOF
and momentum $p$. In the FS, the RICH cannot directly identify protons
with momentum $p<15$ GeV/c. Most of the protons in the settings used
here ($y \sim 2.9$) are below that threshold. Instead, the RICH is
used to veto pions and kaons. The PID and data selection is described
more thoroughly in~\cite{bratio02}.

For each magnet and angle setting the geometrical acceptance is
calculated using a Monte Carlo simulation of the BRAHMS detector. The
corrections for multiple scattering and absorption have been
calculated in the same simulation framework.  Tracking efficiencies
for each chamber have been calculated from the data by comparing the
number of identified track segments in the chamber to the number of
reference tracks determined by other detectors disregarding the
chamber under consideration. For the spectrometers the total tracking
efficiencies are $\sim 92$\% in the MRS and $\sim 60$\% in the FS in
the settings used here.  After all the corrections have been applied
to the data the differential yields are finally projected to the
$m_T$-axis in a narrow rapidity range.  Figure~\ref{fig:spectra} shows
transverse mass spectra for $p$ and $\bar{p}$ at $y=0.0, 0.8, 2.9$.

To find the yields $dN/dy$, the data is fitted with an exponential
function (Eq.~\ref{eq:fitfunc}). The fit is used to extrapolate the
yields to the regions where there is no data (mainly low $m_T$).

\begin{equation}
  \label{eq:fitfunc}
  \frac{1}{N_{ev}}\frac{d^2N}{2\pi m_T dy dm_T} = 
  \frac{1}{2\pi}\frac{dN}{dy}\frac{1}{T(T + m_p)}
  \exp\left(-\frac{m_T-m_p}T\right),
\end{equation}

\noindent where $m_p$ is the proton mass, $m_T = \sqrt{p_T^2 + m_p^2}$ is the
transverse mass, and $T$ is the effective temperature (inverse slope).

\begin{figure}[htb]
  \vspace{-5mm}
  \begin{minipage}[t]{0.45\columnwidth}
    \includegraphics[keepaspectratio, width=0.9\columnwidth]
    {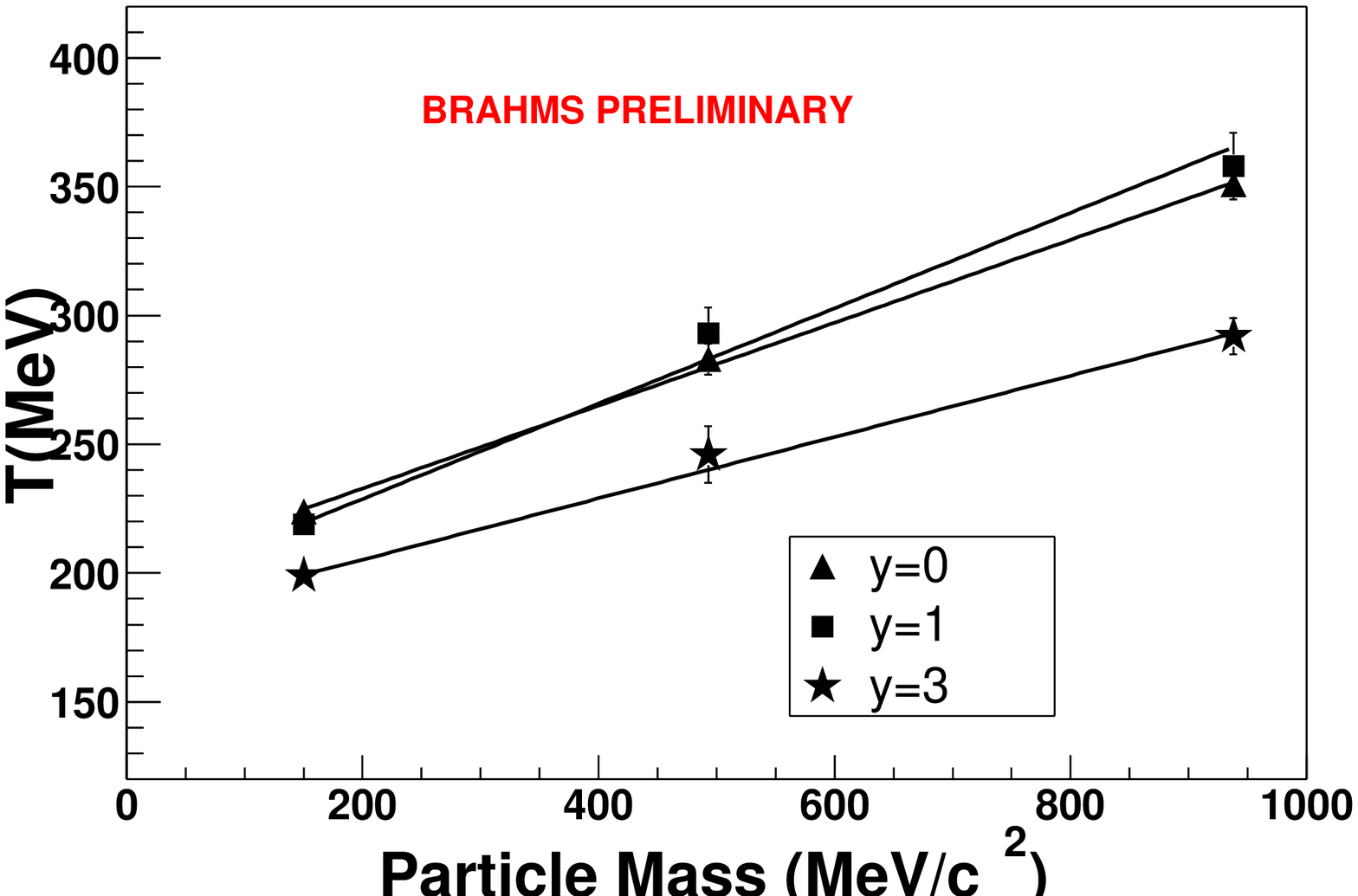}
    \vspace{-5mm}
    \caption{Effective temperatures obtained from fits to pion, kaon,
      and proton spectra. The solid lines connect data points from the
      same rapidity.}
    \label{fig:temperature}
  \end{minipage}
  \hspace{\fill}
  \begin{minipage}[t]{0.5\columnwidth}
    \includegraphics[keepaspectratio, width=1.0\columnwidth]
    {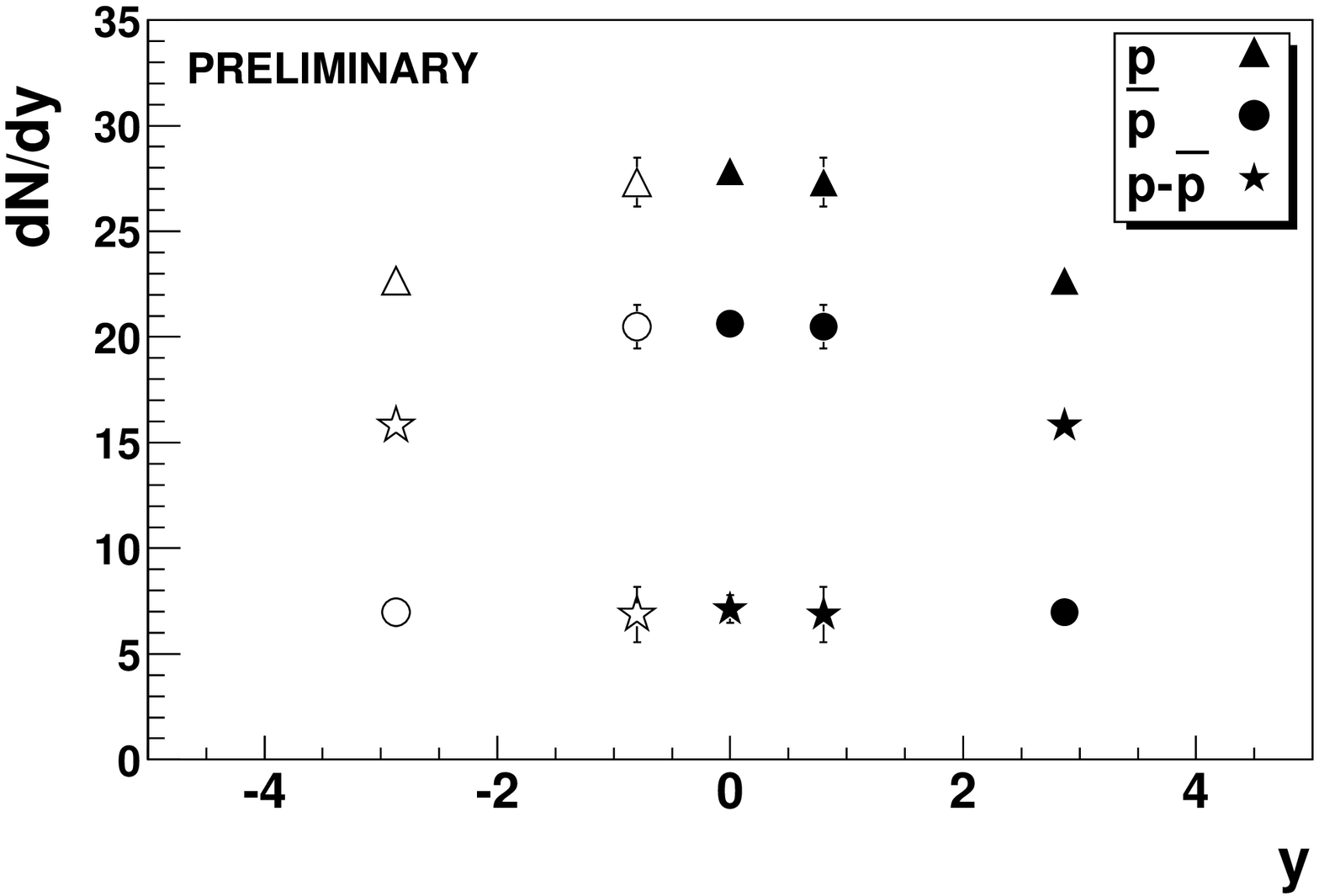}
    \vspace{-10mm}
    \caption{Proton, anti--proton, and net--proton yields as a
      function of rapidity. Event centrality is 0-10\%. Error bars are
      statistical only.}
    \label{fig:yields}
  \end{minipage}
  \vspace{-5mm}
\end{figure}

%
The fit to the data are shown in Fig.~\ref{fig:spectra}. The
$\chi^2/DOF$ is $\sim$ 3--4 for the mid--rapidity data, and $\sim$
0.75--1.6 at the two forward rapidities. The effective temperature
from the fits are the same for protons and anti--protons within 30
MeV.  Fig.~\ref{fig:temperature} shows the rapidity dependence of the
effective temperature for positive pions, kaons, and protons (pions
and kaons are from~\cite{jordre02}). The mass dependence of the
effective temperature is due to strong transverse flow which has a
stronger effect on heavier particles. The fact that the effective
temperature drops moderately as a function of rapidity indicates that
there is strong longitudinal flow, whereas in a thermal isotropic
fireball picture the effective temperature drops as $T(y) =
T(y=0)/\cosh{(y)}$. This would decrease the effective temperature by
90\% from $y=0$ to $y=2.9$ compared with $\sim 10-20$\% observed. The
effective temperature obtained from the fit is very sensitive to the
fit--range used and have a systematic error of 10--15\%.


The rapidity dependence of the extrapolated yields, $dN/dy$, for
protons, anti--protons, and net--protons are presented in
Fig.~\ref{fig:yields}. The width of the proton distribution is wider
than the anti--proton distribution and decreases less than 20\% from
mid--rapidity to rapidity $y=3$, whereas pions decrease by
50\%~\cite{jordre02}. The net--proton distribution increases at
forward rapidities where the original baryon content is recovered.

The current systematic errors on the yields are estimated to be 15\%
in the MRS ($y < 1$) and 20\% in the FS. The largest contribution
comes from the extrapolation to the yield $dN/dy$. This effect should
not change the shape of the distribution, only the overall level. The
systematic error from point to point in rapidity is 10\%.

\section{DISCUSSION}
\label{sec:discussion}

The yields of pions, kaons, and protons drop less than 10\% over the
first unit of rapidity. Because of the systematic error, it is
premature to rule out or confirm that there is a boost invariant
plateau of width $\Delta y \sim$ 2--3 rapidity units around
mid--rapidity. From the charge conjugate ratios~\cite{bratio02} it is
clear that it cannot be broader.

Assuming that the anti--proton yields decrease smoothly between
rapidity 1 and 3, we can deduce from measurements of the rapidity
dependence of $N(\bar{p})/N(p)$ ratio~\cite{bratio02} that the proton
distribution behaves smoothly in the same interval, so the net--proton
distribution does not peak for $y < 3$ i.e., the bulk part of the
net--protons is found far from mid--rapidity.

\begin{figure}[htb]
  \vspace{-5mm}
  \begin{center}
    \includegraphics[keepaspectratio, width=0.65\columnwidth]
    {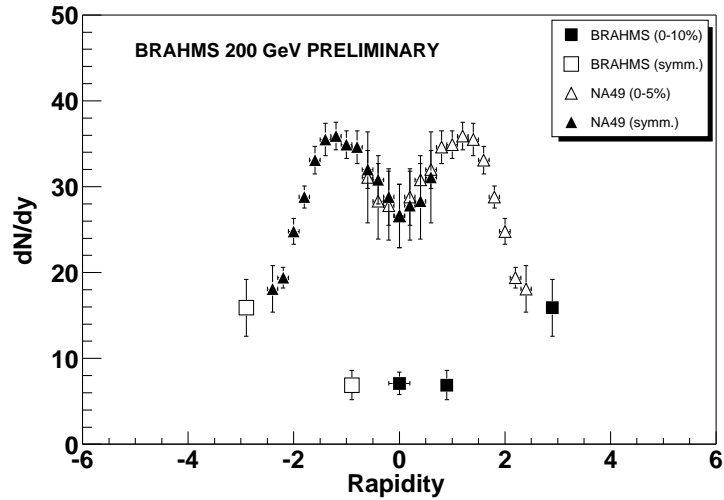}
    \vspace{-10mm}
    \caption{Comparison between net--protons measured at RHIC and
      SPS~\cite{na49}. Note that at SPS, $y_{beam} \sim 2.9$
      whereas at RHIC $y_{beam} \sim 5.4$. The error bars are
      statistical and systematic.}
    \label{fig:sps} 
  \end{center}
  \vspace{-5mm}
\end{figure}


When compared to SPS data (\rootsps) in Fig.~\ref{fig:sps}, it is
clear that the net--proton yield at RHIC is significantly lower
(60--70\%).  At SPS most of the protons at mid--rapidity are
net--protons, while at RHIC the pair production mechanism dominates
the proton production in the central region and the bulk of the
net--protons are found at forward rapidities.

The BRAHMS results shows that the longitudinal flow in the collisions
is strong. The proton and anti--proton yields exhibit little variation
over at least the 2 central units of rapidity. The BRAHMS data suggest
the onset of a boost invariant like plateau around mid--rapidity. The
net--proton content of the central rapidity region is much lower than
what was observed at lower beam energies, indicating that at RHIC
collisions are highly transparent.

\end{document}